\documentclass[prx,twocolumn,amsmath,amssymb,floatfix,footinbib,bibnotes,superscriptaddress]{revtex4-1}
\usepackage{bm}
\usepackage{epsf}
\usepackage{amssymb}
\usepackage{amsmath}
\usepackage{graphicx}
\usepackage{rotating}
\usepackage{epsfig}
\usepackage{psfrag}
\usepackage{amsmath}
\usepackage{hyperref}
\usepackage{subfigure}
\usepackage{bm}
\usepackage[usenames,dvipsnames]{color}

\renewcommand{\a}{\alpha}

\renewcommand{\d}{\delta}

\newcommand{\nn}{\nonumber}

\def \ket#1{{\,|\,#1\,\rangle\,}}
\def \bra#1{{\,\langle\,#1\,|\,}}
\def \bea{\begin{eqnarray}}
\def \eea{\end{eqnarray}}

\newcommand{\be}{\begin{equation}}
\newcommand{\ee}{\end{equation}}
\newcommand{\ba}{\begin{eqnarray}}
\newcommand{\ea}{\end{eqnarray}}

\newcommand {\apgt} {\ {\raise-.5ex\hbox{$\buildrel>\over\sim$}}\ }
\newcommand {\aplt} {\ {\raise-.5ex\hbox{$\buildrel<\over\sim$}}\ }

\begin{document}

\title{Quantum thermalization dynamics with Matrix-Product States}

\author{Eyal Leviatan}
\affiliation{Department of Condensed Matter Physics, Weizmann Institute of Science, Rehovot 7610001, Israel}

\author{Frank Pollmann}
\affiliation{Max-Planck-Institut f\"ur Physik komplexer Systeme, N\"othnitzer Str. 38, 01187 Dresden, Germany}
\affiliation{Technische Universit\"at M\"unchen, Physics Department T42, 85747 Garching, Germany}

\author{Jens H. Bardarson}
\affiliation{Max-Planck-Institut f\"ur Physik komplexer Systeme, N\"othnitzer Str. 38, 01187 Dresden, Germany}
\affiliation{Department of Physics, KTH Royal Institute of Technology, Stockholm SE-10691, Sweden}

\author{David A. Huse}
\affiliation{Physics Department, Princeton University, Princeton, NJ 08544, USA}

\author{Ehud Altman}
\affiliation{Department of Physics, University of California, Berkeley, CA 94720, USA}

\begin{abstract}
We study the dynamics of thermalization following a quantum quench using tensor-network methods. Contrary to the common belief that the rapid growth of entanglement and the resulting exponential growth of the bond dimension restricts simulations to short times, we demonstrate that the long time limit of local observables can be well captured using the time-dependent variational principle. This allows to extract transport coefficients such as the energy diffusion constant from simulations with rather small bond dimensions. We further study the characteristic of the chaotic wave that precedes the emergence of hydrodynamics, to find a ballistic diffusively-broadening wave-front.
\end{abstract}

\maketitle
\section{Introduction}
The question how quantum systems thermalize, or fail to do so, when evolving under their own intrinsic dynamics is an important challenge for theoretical physics, which recently came within the reach of experiments in ultra-cold atomic systems \cite{sadler2006,kinoshita2006,strohmaier2010,gring2012,schreiber2015,luschen2016}.
Recent studies of holographic models \cite{kovtun2005,hartnoll2015,Maldacena2015,Mezei2017} as well as certain exactly solvable strongly coupled field theories, such as the Sachdev-Ye-Kitaev model \cite{Sachdev1993,KitaevKITP,Maldacena2016,Gu2016,davison2016}, have uncovered a wealth of emergent structure in the dynamics of thermalizing systems that has not been appreciated before and is beyond the semiclassical Boltzmann description of thermalization. For example, these studies have suggested the existence of fundamental quantum bounds on transport, thermalization rates and chaotic dynamics. These new results also lead to conjecture possible universal relations between the characteristics of chaos, which determine the rate of information scrambling in a system, and hydrodynamic transport coefficients, which control the long time relaxation of physical observables \cite{Blake2016}. 

The new ideas on quantum thermalization need to be tested with generic physical models and this, in turn, requires new methods for computing quantum time evolution. Almost all the numerical calculations done so far to address this problem relied on exact diagonalization  (see, e.g., Refs.~\cite{Oganesyan2007,rigol2008,barlev2015,agarwal2015,fu2016}), which is severely constrained by the attainable system sizes. On the other hand, the density matrix renormalization group (DMRG) \cite{W92} and other tensor network based methods can tackle large systems, but are considered inadequate for computing the long time dynamics. The main obstruction is the rapid growth of entanglement entropy, which translates to exponential growth with time of the tensor (bond) dimensions needed for the calculation \cite{Prosen2007}. 

In this paper, we propose a way to overcome the limitation on computation of  long time dynamics within the manifold of tensor networks. A first clue that something is amiss in the usual assessment of the entanglement growth as a fundamental obstruction comes from the observation that quantum thermalizing systems are governed at long times by emergent classical hydrodynamics. Given the universality of the hydrodynamic behavior, dictated by conservation laws and a small number of parameters, we do not expect  the dynamics of local observables to depend  on the detailed long range entanglement structure of the micro-state.  The pertinent question then is how to truncate the growing entanglement entropy without sacrificing crucial information on local observables?

Previous attempts to restrain the growth of entanglement entropy involved the use of matrix-product operators (MPO) \cite{banuls2009, Zanadi2001, Prosen2007, Dubail2017} or purified states \cite{Feiguin05, Barthel09, karrasch2012, karrasch2013} to represent density matrices. These works sought to utilize an inherent advantage in using thermal density matrices rather than pure states: while in pure states the thermodynamic entropy is encoded entirely through long range quantum correlations (entanglement entropy), in a density matrix it can be represented as a statistical mixture of local degrees of freedom (or local entanglement with ancilla bits in a purified state). Thus, for example, the infinite temperature density matrix is simply a unit matrix, equivalent to a direct product of unit matrices representing the local mixed states. When perturbed by a local operator (e.g. by imposing a spin up at the origin), such a system is expected to relax back to effective local equilibrium over a short timescale set by the local interactions. One might hope that the local equilibrium state attained in the process can again be described in terms of an MPO (or purified state) with short range quantum correlations, similar to the initial state. Unfortunately, the exact time evolution leads to a linear growth of the entanglement entropy of the density matrix in time. Attempts to curb the entanglement growth, while still capturing the exact density matrix, have met with only partial success, allowing to reduce the growth rate somewhat \cite{banuls2009,karrasch2012,karrasch2013}. But the fundamental problem of an exponentially growing bond dimension remains.  

Here we take a different approach, which aims to truncate the entanglement growth within a systematic approximation, rather than attempting to capture the exact dynamics.  To this end we employ the time dependent variational principle (TDVP) \cite{dirac1930,jackiw1979}  to time-evolve matrix product states (MPS) \cite{Fannes-1992} within a space of fixed bond dimension $\chi$, using an efficient algorithm proposed by Haegemann {\it et al.} \cite{Haegeman2011,Haegeman2016} (see also \cite{Ueda2006,Kinder2011}). The entanglement entropy in this approach is capped by $\log \chi$. Hence, the MPS is not even a nearly approximate description of the micro-state, which would naturally evolve to volume law entanglement entropy. In the language of time dependent MPS calculations, the truncation error is bound to become large after a short time.

Why then should the TDVP scheme nonetheless capture the long time dynamics of thermalizing systems? A crucial feature for our purpose is that the TDVP respects conservation laws regardless of the truncation. This is in contrast to the common time dependent DMRG methods (such as the time evolving block decimation (TEBD) \cite{Vidal2003}) which violate them when the truncation error becomes large. Indeed the TDVP generates nonlinear classical dynamics in the variational manifold, driven by a classical Hamiltonian having the same symmetries as the original quantum Hamiltonian. Because such nonlinear dynamics is generically chaotic, the hydrodynamic behavior of local observables is guaranteed to emerge at long times even if we only keep a small bond dimension. Of course, we are not guaranteed a priori that the hydrodynamics in this scheme is governed by the correct transport coefficients. On physical grounds, however, we expect that these are determined by quantum processes that occur  on rather short scales related to the short thermalization time and possibly the thermal coherence length. Such processes can in principle be captured by MPS with finite bond dimension. Increasing the bond dimension of the variational family of states allows to systematically improve the calculation and to assess the accuracy of the result by checking for convergence with $\chi$.

To test the new approach, we consider the dynamics of the Ising chain with both longitudinal and transverse fields:
\be
H = J\sum_{i=0}^{N-1}{S^x_i S^x_{i+1}} + h_x\sum_{i=0}^N{S^x_i} + h_z\sum_{i=0}^N{S^z_i} \label{Model}.
\ee
Here $S^\a_i$ are spin-$1/2$ operators defined on site $i$. This model, in a regime of parameters far from any integrable point, is commonly used as a testbed for thermalization dynamics \cite{hastings2011,Roberts2014}. A simplifying feature is the low symmetry, which leaves energy as the only conserved quantity. In our calculations $N=100$ (i.e., the chain length is 101 sites) and we use the coupling constants $\{J,h_x,h_z\}=\{1,0.25,-0.525\}$, for which the system is indeed far from any integrable point and shows fast thermalization \cite{hastings2011}.

We use the TDVP to compute the dynamics induced by the application of a local perturbation to a thermal ensemble of initial states, as explained in more detailed below. We compute two types of quantities: First, we look at relaxation of local observables following the quench to find the expected long time tail associated with energy diffusion. From this we extract the energy diffusion coefficient, which shows excellent convergence with increasing bond dimension $\chi$. Second, we compute a diagnostic of chaos closely related to out-of-time-order correlations. We find a chaotic front that propagates ballistically in the system, but broadens in time diffusively. This behavior is in agreement with results obtained within a model of spin chains evolved with random local unitaries \cite{VonKeyserlingk2017,Nahum2017}.

\begin{figure}[t]
\begin{center}
\includegraphics[width=0.45\textwidth]{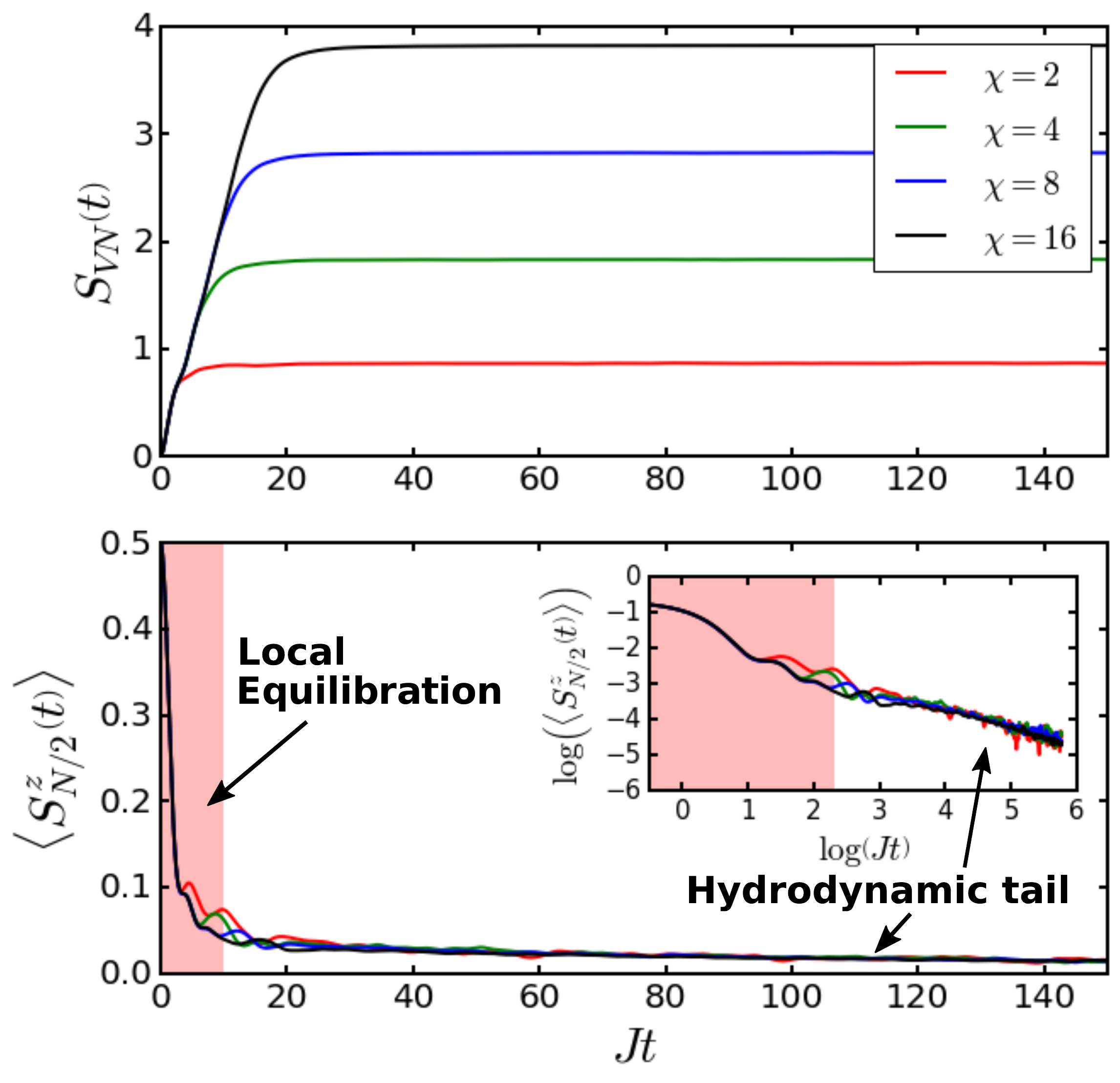}
\llap{\parbox[c]{13.5cm}{\vspace{-14.5cm}\textbf{(a)}}}
\llap{\parbox[c]{13.5cm}{\vspace{-7.15cm}\textbf{(b)}}}
\caption{(a) The entanglement entropy $S_{VN}$ of a bi-partition at site $N/2$ following a quantum quench to an initial ensemble of random product states. As expected the entropy increases linearly in time and saturates to a value roughly equal to $\log_2$ of the bond dimension $\chi$. (b) Relaxation dynamics of the perturbed middle spin in the same calculation. A fast decay which leads to local equilibration is seen at short times followed by a slow hydrodynamic (diffusive) tail decaying as $1/\sqrt{t}$. The inset shows the decay on a log-log scale emphasizing the long time tail.}  
\label{fig.entropy}
\end{center}
\end{figure}

\section{Method}
We now describe the application of the TDVP approach to the problem in some more detail. As mentioned, the TDVP imposes classical dynamics in a phase space defined by the parameters of a variational state $\ket{\psi\left[\alpha\right]}$ through the effective Lagrangian:
\begin{equation}\label{Lvar}
\mathcal{L}\left[\alpha,\dot\alpha\right]=\bra{\psi\left[\alpha\right]}i\partial_t\ket{\psi\left[\alpha\right]}-\bra{\psi\left[\alpha\right]}H\ket{\psi\left[\alpha\right]}.
\end{equation}
In our context the variational manifold is the space of MPS with fixed bond dimension $\chi$:
\begin{equation}
\ket{\psi\left[\alpha\right]}=\sum_{\sigma_0\cdots\sigma_N} {A^0_{\sigma_0}\cdots A^N_{\sigma_N} \ket{\sigma_0\cdots\sigma_N}}.
\end{equation}
The variational time evolution is implemented in each time step $\delta t$ through the application of effective {\em single site} evolution operators on all the matrices $A^0_{\sigma_0}$ to $A^N_{\sigma_N}$ in succession, as described in Ref. \cite{Haegeman2016}. The effective single site evolution operator is obtained from an effective single site Hamiltonian generated by contracting, from both sides, the underlying Hamiltonian with the truncated MPS having the matrix corresponding to that site removed. 

Our goal is to compute the evolution of a local perturbation applied to a thermal ensemble. Thus, we compute time dependent quantities following a quench as averages over an ensemble of initial states. The initial states are chosen to represent the suitable canonical ensemble, perturbed by the application of a local operator on the middle of the chain. 

As a first demonstration of the method we start from a sample of random product states with the direction of the spin on each site chosen independently from a uniform distribution on the Bloch sphere. At this point the ensemble represents the infinite temperature state. The ensemble is quenched by the application of the single site operator $S^+_{N/2}$ on the middle site of the chain (followed by normalization) for every state in the ensemble. The states of the quenched ensemble are then evolved using the TDVP scheme with a given fixed bond dimension $\chi$. 

Fig.~\ref{fig.entropy}(a) shows the average Von-Neumann entanglement entropy for a bi-partition of the evolved states at site $N/2$. Here we define the Von-Neumann entanglement entropy of a subregion $A$ to be $S_{VN}=\text{tr}\rho_A\log_2\rho_A$. This half-system entanglement entropy is seen to grow linearly in time and saturate to a value proportional to the bond dimension $S_{VN}(\infty)\sim  \log_2 \chi$ (upto a small shift). Hence the curves corresponding to different values of $\chi$ depart strongly from each other. In contrast the relaxation of the local operator $S^z_{N/2}$ on the quenched site, seen in Fig.~\ref{fig.entropy}(b), has quantitatively the same behavior for the different values of $\chi$. 

We see two stages in the relaxation of $S^z_{N/2}$. First is a rapid decay of the expectation value as the system reaches local thermal equilibrium in a few microscopic time units. The second stage, seen more clearly in the inset of Fig.~\ref{fig.entropy}(b) is a power law decay of the residual spin expectation value. This is the expected hydrodynamic long time tail. Note that $S^z$ is not conserved, the hydrodynamic tail arises because this operator has a finite overlap with the energy density; it is a manifestation of energy diffusion in the system. This is a first demonstration that such long time thermalization dynamics can be captured accurately using low entanglement states.

For a more accurate study of the thermalization dynamics in what follows we use a different ensemble of initial states. Instead of taking a sample of random product states we consider fully random MPS of fixed bond dimension $\chi$ right at the outset. That is, we do not restrict the initial states to be product states. This has the advantage that each state of this ensemble is in itself more representative of the infinite temperature state as it typically corresponds to the maximal entanglement entropy for the given $\chi$. This facilitates faster convergence with sample size. As before, the quench operator $S^+_{N/2}$ is applied to the middle site for every state (followed by normalization) to produce the ensemble of initial states. 

Before proceeding we note that it is possible to similarly study a quench of a finite temperature state. This is done by evolving the random MPS in imaginary time (and normalizing the state) before applying the quench operator. In this paper, however, we we focus on thermalization and chaotic dynamics at infinite temperature.

\begin{figure}[t] 
\includegraphics[width=0.45\textwidth]{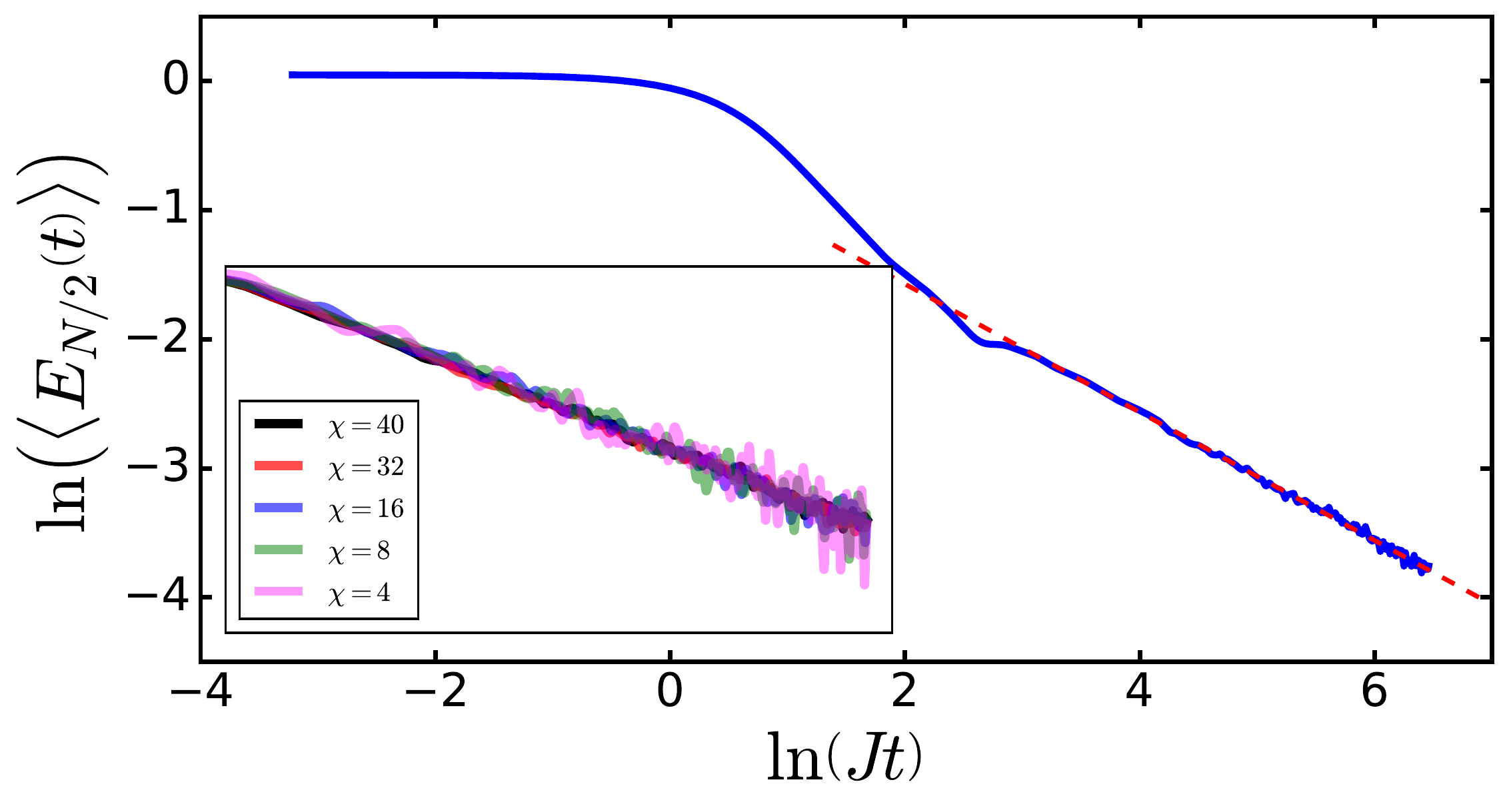}
\llap{\parbox[c]{13.5cm}{\vspace{-7.75cm}\textbf{(a)}}}
\includegraphics[width=0.45\textwidth]{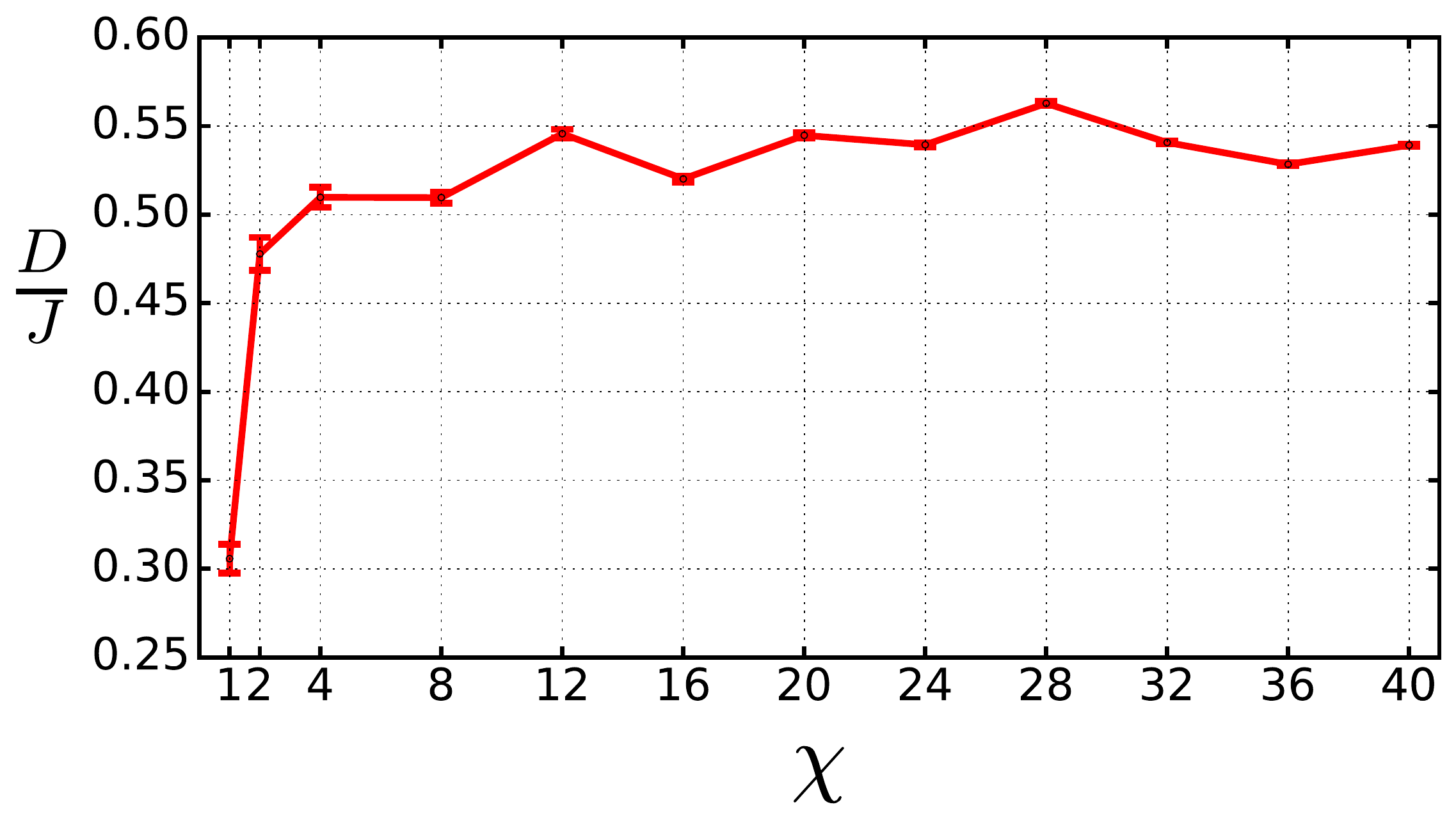}
\llap{\parbox[c]{13.5cm}{\vspace{-7.75cm}\textbf{(b)}}}
\caption{(a) Decay of the energy on the middle site of the chain after the quench agrees perfectly with a diffusive decay as $\sim1/\sqrt{t}$. This plot corresponds to a 101 site chain and bond dimension $\chi=32$. Similar behavior is seen for different $\chi$. (b) The diffusion constant extracted from the energy relaxation computed with different bond dimensions shows convergence beyond $\chi\approx 4$.} 
\label{fig.ExampleSim}
\end{figure}

\section{Energy Diffusion}
The diffusive tail observed in the relaxation of a local observable emerges because that observable has a finite overlap with the local energy. The perturbation $S^+_{N/2}$ applied to the thermal state has injected energy locally to the middle of the chain. Because of energy conservation, this excess energy is expected to spread out diffusively away from the perturbed site at long times with a profile:
\be
E\left(x,t\right)=\frac{E_0}{\sqrt{2\pi D t}}\exp\left[ -{x^2 \over 2 D t}\right],\label{Diffusion}
\ee
where $x$ is the distance from the middle site. This is indeed the behavior found in the calculations, which show, for all bond-dimensions, a decay of the energy of the middle site $\left(x=0\right)$ as $1/\sqrt{t}$. Fig.~\ref{fig.ExampleSim}(a) depicts an example of the energy relaxation on the middle site, obtained from averaging over $1000$ random initial MPS of bond-dimension $\chi=32$. Note the separation of time scales between a fast decay to local equilibrium, which happens over a time of order $J^{-1}$ and the slow hydrodynamic decay that sets in afterward. The diffusion constant extracted from the fit to Eq.~(\ref{Diffusion}) for $x=0$ is seen in Fig.~\ref{fig.ExampleSim}(b) for bond dimensions $1$ through $40$. The result shows rapid convergence with bond dimension for $\chi\ge4$. Note that we gradually reduced the number of initial states averaged over as we increased the bond dimension. Specifically we used  6000 initial states for $\chi=1,2$ and gradually went down to 500 initial states for $\chi=36,40$, where the calculations are most demanding. We note, however, that at high bond dimension the state to state fluctuations are much smaller and thus they require less averaging. 

The errors in fitting the diffusion constant are estimated by the following procedure. We add simulated Gaussian noise to the average curve of $E(x=0,t)$ with an amplitude determined by the variance of the fluctuations in this quantity (over the distribution of initial states) and a correlation time of $10/J$. We fit the diffusion constant from an ensemble of such simulated noisy curves in order to assess the fit error.

\section{Quantum chaos} 
The emergence of hydrodynamic transport is closely linked to quantum chaos. Hydrodynamics can ensue only if the dynamics becomes effectively irreversible through the scrambling of information by chaos. Below we characterize the chaotic dynamics in our model. 

As a first step we define a diagnostic of chaos which is convenient to compute with MPS. Classical chaos is characterized by exponential divergence of nearby phase space trajectories. Quantum wavefunctions cannot diverge from each other in the same way because the unitary evolution preserves distance in Hilbert space. Instead, let us consider states of a subsystem, described by the appropriate reduced density matrices. A measure of distance between these two reduced density matrices can diverge when the system as a whole is undergoing quantum unitary evolution similarly to the divergence of classical trajectories. 
\begin{figure}[t]
\includegraphics[width=0.45\textwidth]{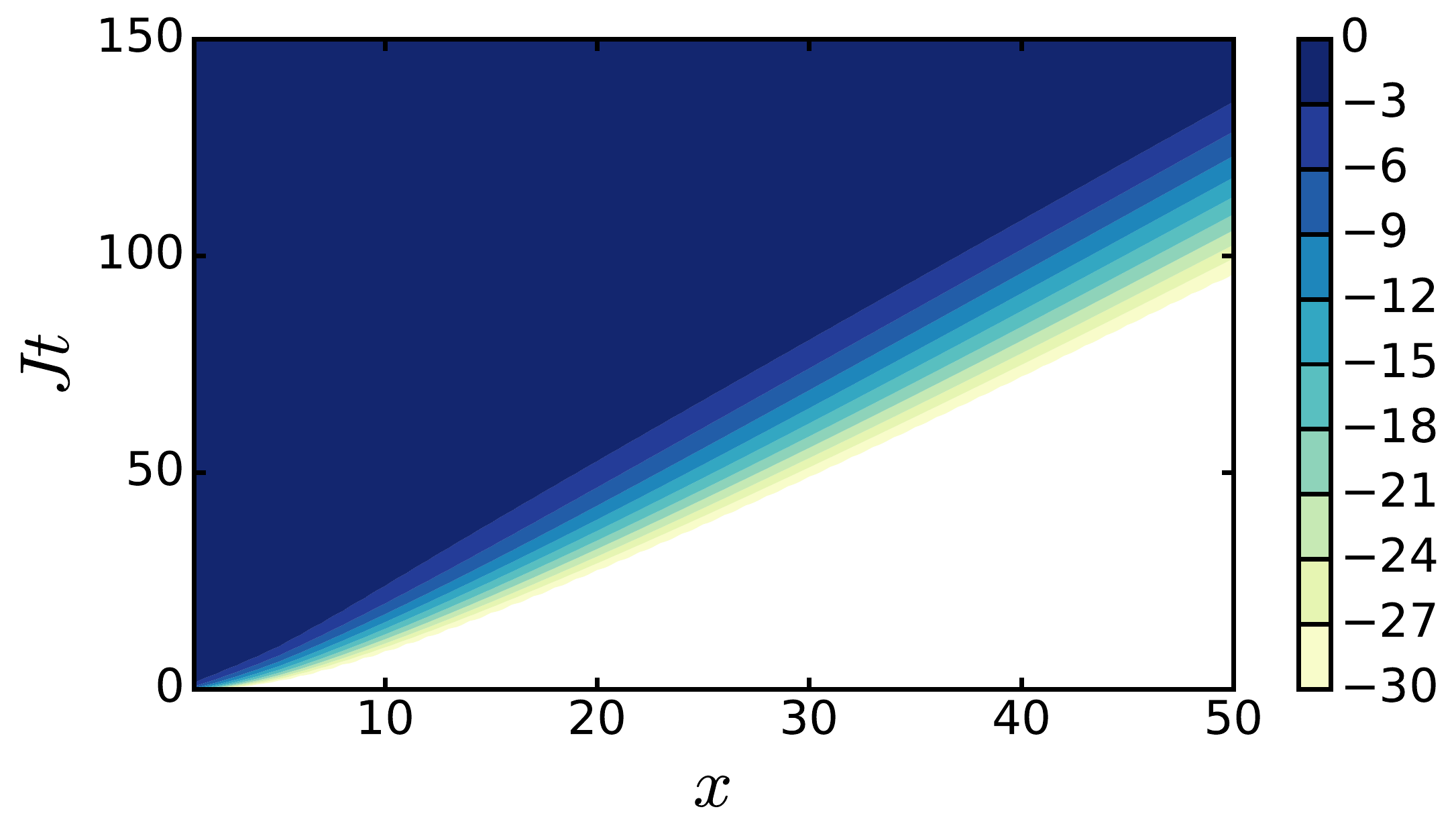}
\caption{Logarithm of the normalized average distance measure $\delta^2$ as a function of space and time, obtained for $\chi=32$. The colormap shows a ballistic propagation of the chaotic wave with a velocity $v_B$ known as the ``butterfly velocity''. This is in contrast to the diffusive energy transport in the system. Also visible in the colormap is a broadening of the chaotic wave front. Further analysis shows this broadening to be diffusive (see Fig.~\ref{fig.chaos_x} below).} 
\label{fig.chaosColor}
\end{figure}

For concreteness consider the spin model (\ref{Model}). For each state in the ensemble of initial conditions we compare the time-evolution of an unperturbed state denoted by $|\psi_1\rangle$ with that of a state perturbed by a local unitary operator on the left edge (first site) of the chain $|\psi_2\rangle=S^x_0|\psi_1\rangle$. Note that because the perturbation is unitary the two initial states $|\psi_1\rangle$ and $|\psi_2\rangle$ differ only locally. The perturbation does not alter correlations away from the perturbation site. The same would not be true if the perturbation was the nonunitary operator $S^+$ we applied in the previous section to investigate the relaxation of local observables. The nonunitary operator immediately gives rise to a non local effect, which decays exponentially with a correlation length of order $\log\chi$ away from the perturbation. 

Both the perturbed and unperturbed copies of the initial state are  evolved using the TDVP scheme. To assess the difference between these states we make bi-partitions of the chain at bonds located at increasing distances, $x$, from the left most bond. We then consider the reduced density matrices of the partition to the right of the bond at $x$, $\rho_1^R(x,t)$ and $\rho^R_2(x,t)$, corresponding to the unperturbed and perturbed states respectively. As a measure of the distance between the two states we use the Frobenius norm of the difference between them:
\bea
d^2\left(x,t\right)&\equiv& {\text{tr}}\left(\left[\rho_1^R\left(x,t\right)- \rho_2^R\left(x,t\right)\right]^2\right)
\label{d2}
\eea

It is important that $x<L/2$ to ensure that $\rho^R_{1/2}$ represent the reduced states covering more than half the system. Reduced states corresponding to less than half the system are expected to approach a thermal density matrix independent of the initial condition. Hence such states do not significantly depart from each other (see appendix). 

The distance measure $d^2$ depends on the dimension of the effective Hilbert space to the right of the position $x$. In the TDVP scheme this dimension is bounded by the bond dimension used in the MPS description of the state. In order to draw comparisons between different bond dimension, we present results for a normalized version of the distance measure:
\bea
\d ^2\left(x,t\right) = \frac{d^2\left(x,t\right)}{d^2\left(x,\infty \right)}
\label{delta2}
\eea

We note that the quantity $d^2$ defined above in (\ref{d2}) can be measured by interfering two copies of the time dependent quantum state \cite{Daley2012}, using the experimental set-up developed to measure the second Renyi entropy in Ref. \cite{Islam2015}. For example, to measure the overlap ${\text{tr}}\left(\rho^R_1\rho^R_2\right)$ one would need to perturb only one of the two copies at $t=0$, before interfering them at time $t$ \footnote{We thank Markus Greiner for pointing out this possibility}.

\begin{figure}[t]
\includegraphics[width=0.45\textwidth]{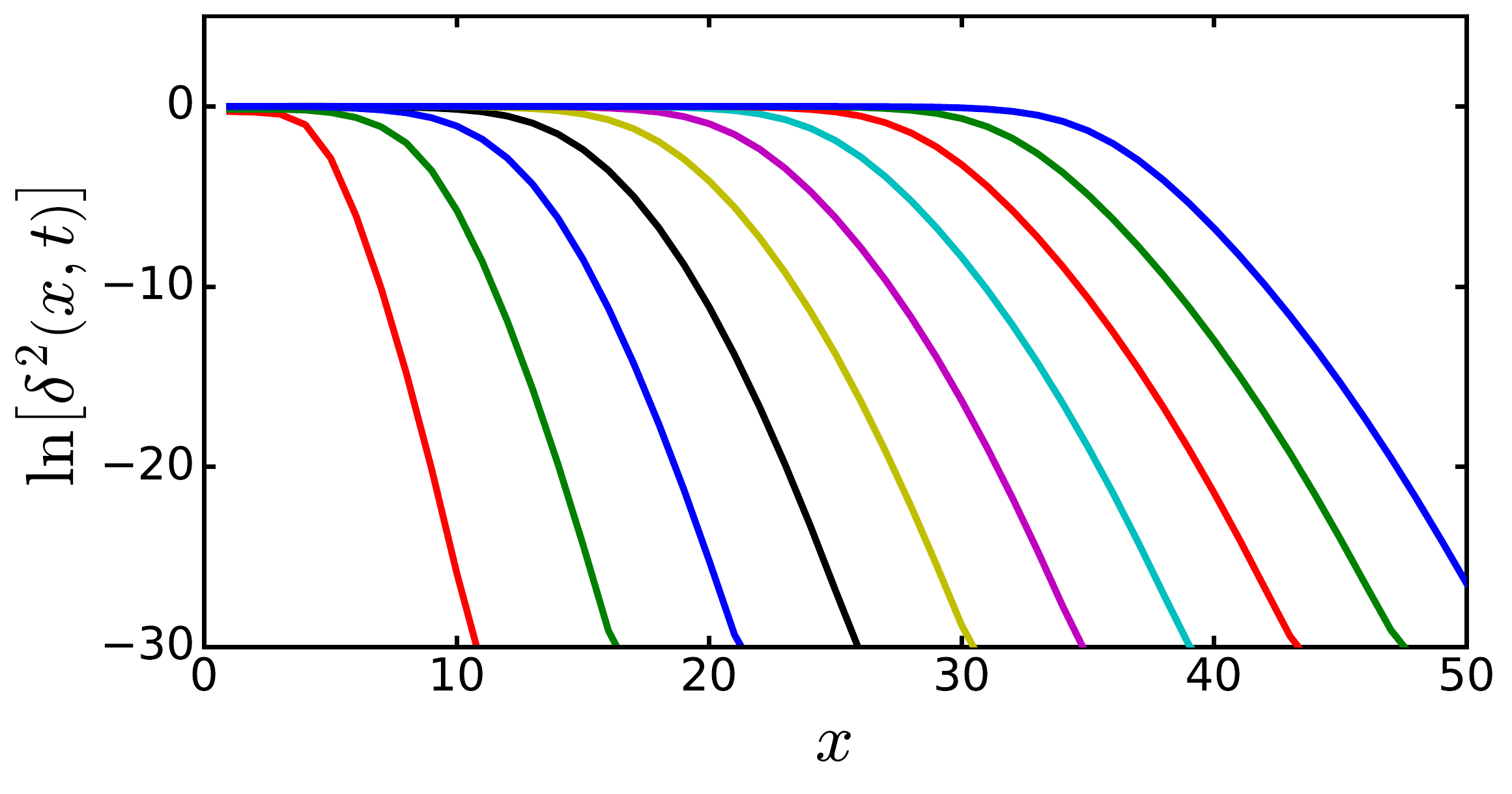}
\llap{\parbox[c]{13cm}{\vspace{-7.6cm}\textbf{(a)}}}
\includegraphics[width=0.45\textwidth]{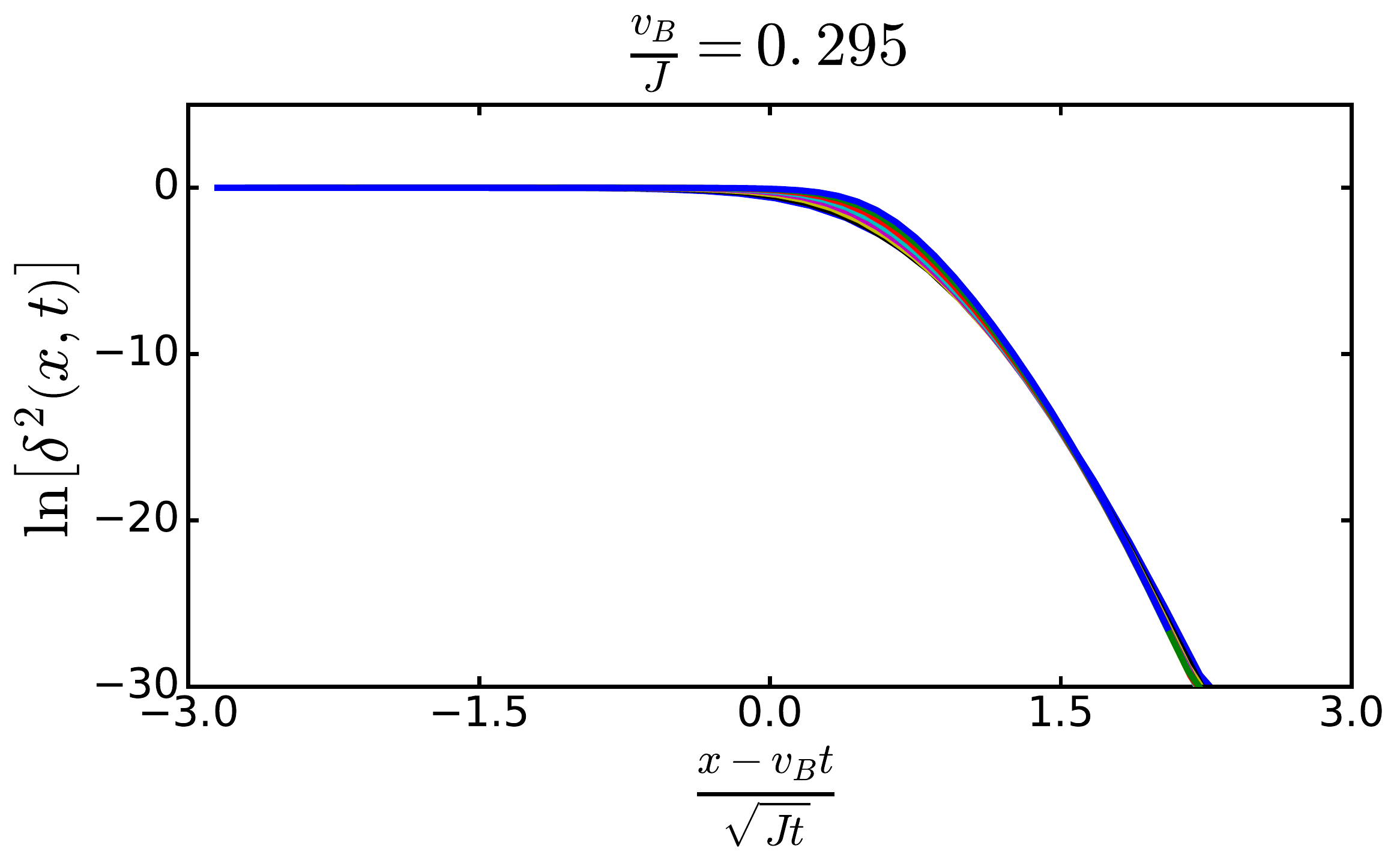}
\llap{\parbox[c]{13cm}{\vspace{-8.1cm}\textbf{(b)}}}
\caption{(a) Logarithm of the normalized average distance measure $\delta^2$ as a function of the distance from the perturbation, at times $10\leq Jt \leq100$ (in steps of $\Delta Jt=10$), obtained for $\chi=32$. (b) Scaling collapse of the $20\leq Jt \leq 100$ curves corresponding to a ballistic diffusively-broadening wave-front: $\left(x-v_Bt\right)/\sqrt{t}$, with $v_B=0.295 J a$.} 
\label{fig.chaos_x}
\end{figure}

\begin{figure}[t]
\includegraphics[width=0.45\textwidth]{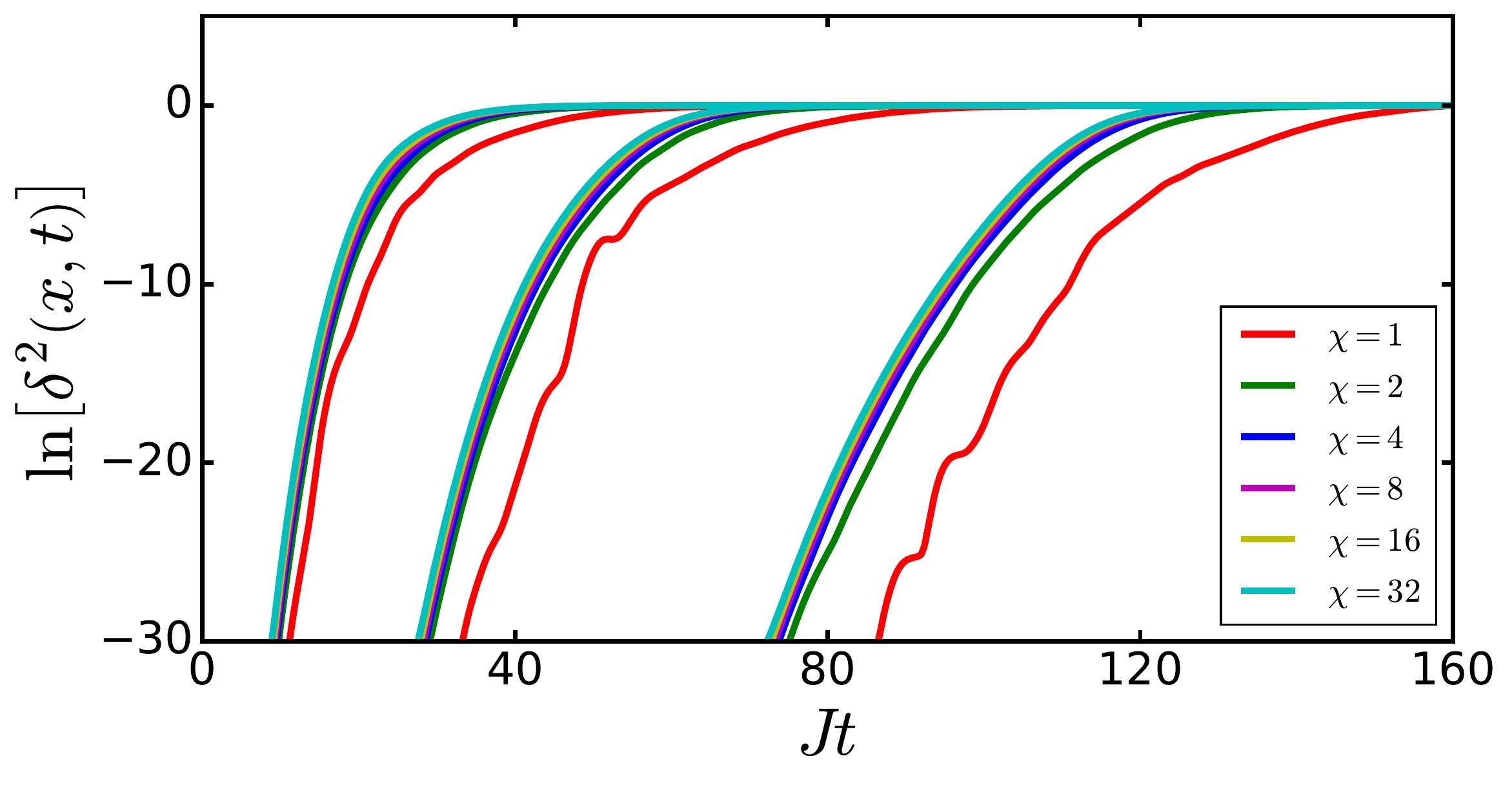}
\llap{\parbox[c]{13.5cm}{\vspace{-7.55cm}\textbf{(a)}}}
\includegraphics[width=0.45\textwidth]{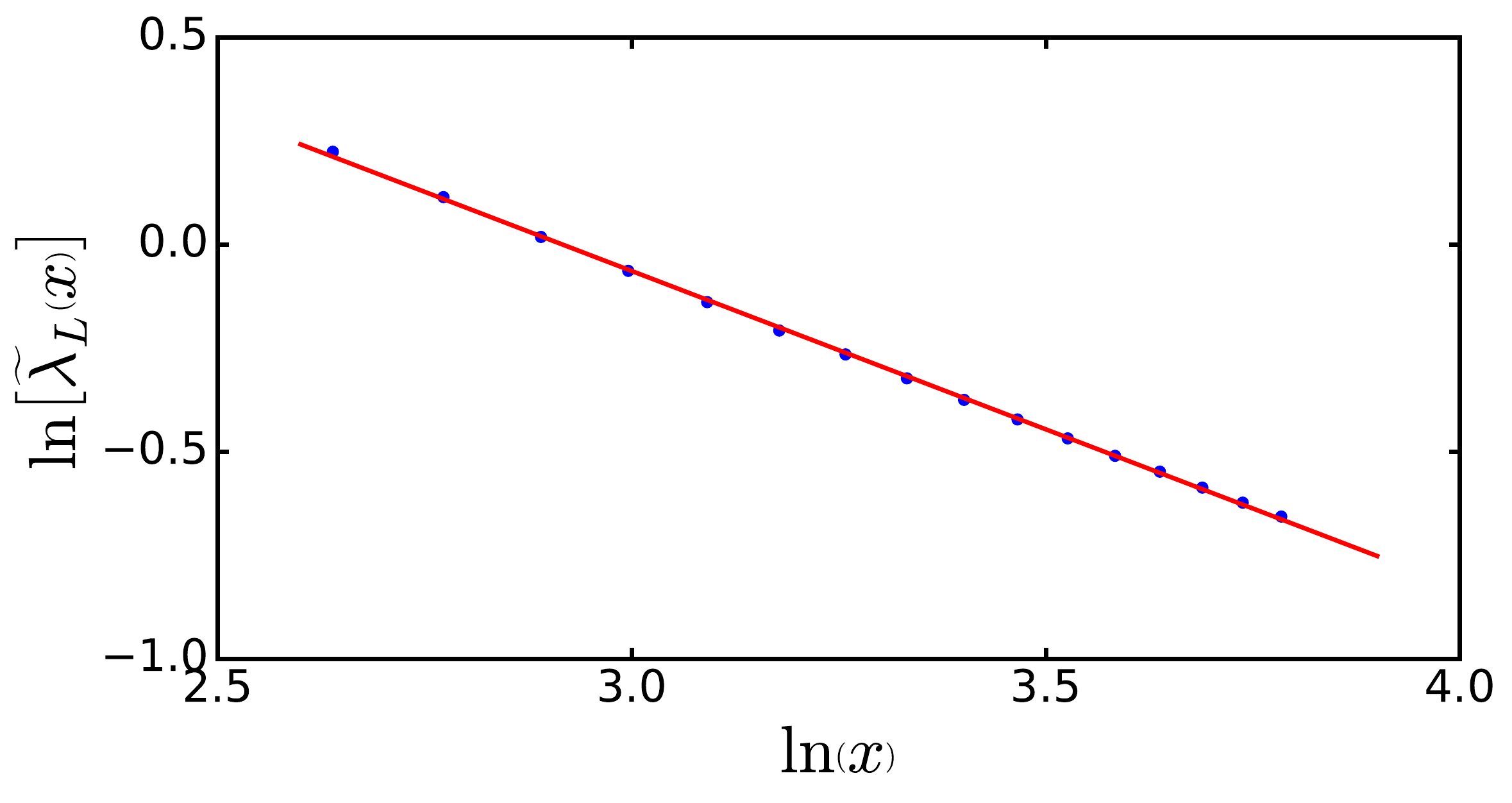}
\llap{\parbox[c]{13.35cm}{\vspace{-7.5cm}\textbf{(b)}}}
\caption{Logarithm of the normalized average distance measure $\delta^2$ as a function of time, at distances $x=10,20,40$ from the perturbation. Here we compare results obtained for bond dimensions: $\chi=1,2,4,8,16,32$. (b) The logarithm of the effective Lyapunov exponent $\widetilde{\lambda}_L$ as a function of the logarithm of the distance $x$ from the perturbation. The linear fit shows a power law decay $1/x^z$ with $z\approx0.76$.} 
\label{fig.chaos_t}
\end{figure}

Fig.~\ref{fig.chaosColor} shows a colormap of $\ln \delta^2(x,t)$ obtained from averaging over $500$ initial states with bond dimension $\chi=32$. The region in which the distance measure grows appears to travel away from the perturbation with a constant velocity
commonly referred to as the ``butterfly velocity". As in the well known classical butterfly effect, the butterfly velocity quantifies how the effect of a small perturbation propagates to cause a big effect on a distant part of the system. In a classical chaotic system one expects the distance $\delta$ between the perturbed and unperturbed states to grow exponentially as $e^{\lambda t}$, where $\lambda$ is the Lyapunov exponent characterizing the chaos. In Fig.~\ref{fig.chaosColor}, the growth front of $\ln \delta$ appears to broaden with the distance $x$ (or time $t$) from the source of the perturbation, which is inconsistent with having a single Lyapunov exponent for the system.

To better characterize the broadening of the front we first analyze constant time snapshots of it, i.e., we consider $\ln \delta^2$ plotted as a function of $x$ at different times as shown in Fig.~\ref{fig.chaos_x}(a). In each plot we see how $\delta^2$ grows from a small value at large distances that the front has not yet reached, to the asymptotic value of one at shorter distances. Rescaling the plots of Fig.~\ref{fig.chaos_x}(a) brings out the self similarity of the growth front and characterizes how it broadens as it propagates. As a first step we shift the curves on the $x$ axis by transforming $x\to x-v_B t$, where $v_B$ is the butterfly velocity. Then, we rescale the $x$ axis by $\sqrt{t}$, which collapses all the curves to a single curve. This scaling collapse implies diffusive broadening of the front as $\Delta x\sim \sqrt{D_B t}$, with $D_B$ a diffusion constant describing the broadening of the chaotic front. The single fitting parameter in this scheme is the butterfly velocity, which we determine to be $v_B=0.295Ja$. Note that it is larger than the velocity implicit in the linear growth of the entanglement entropy $v_E \equiv d S_{VN}/ dt\approx0.24Ja$ shown in Fig.~\ref{fig.entropy}(a). This is in agreement with the predictions of Refs.~\cite{Nahum2016, Mezei2017}, which argued that the entanglement velocity, being constrained by strong sub-additivity, must be smaller than the speed of operator spreading. Note that the butterfly velocity that we measure is expected to approach the speed of operator spreading in the high temperature limit in which we work here \cite{Roberts2014}.

Fig.~\ref{fig.chaos_t}(a) shows the growth of $\delta^2$ with time at different distances $x$ and for different bond dimensions. As also shown in the appendix through comparison to exact results, the calculations appear to be converged already for bond dimension of about $\chi=4$.   Crucially we see that, due to the broadening of the chaotic front, the characteristic growth rate of $\delta(x,t)$ decreases with the distance from the perturbation. Hence it cannot be described by a single Lyapunov exponent. To show this explicitly, we define an effective Lyapunov exponent at a distance $x$ as the ``initial" growth rate of $\delta(x,t)$:  $\widetilde{\lambda}_L(x)\equiv \partial\ln \delta(x,t=t_0)/\partial t$, where $t_0$ is the time at which $\delta^2(x,t_0)$ reaches a threshold value $\delta^2_{th}$ above machine precision. The effective Lyapunov exponent defined with a threshold of $\delta^2_{th}=e^{-25}$ is plotted in Fig. \ref{fig.chaos_t}(b), showing an approximate power-law decrease with the distance $x$.


\section{Discussion}
We have shown that calculations using the time dependent variational principle (TDVP) with matrix-product states (MPS) of low bond dimension can capture the essential long time  dynamics of thermalizing (closed) quantum systems at high temperature. This runs contrary to the conventional wisdom that the rapid growth of entanglement entropy severely limits tensor network calculations at long times. 
Indeed hydrodynamic behavior is naturally obtained in our calculations, allowing to extract transport coefficients such as the energy diffusion constant.  A crucial feature of  TDVP in this context is that it satisfies energy conservation and other conservation rules. For this reason the emergence of hydrodynamic behavior is essentially guaranteed for any bond dimension. The correct transport coefficients and chaos parameters are obtained in spite of the truncation because at high temperature they are generated by quantum processes on a short scale that can be captured by MPS with low bond dimension.  
The accurate results obtained at very low bond dimension (even $\chi=2$) lends hope that this method may also be useful for computing the dynamics of systems in two or three dimensions. 

It would be interesting to apply this method also to study the dynamics at finite temperature, especially in the vicinity of quantum critical points, where the system remains strongly coupled down to the lowest temperature. Calculations in this regime may allow to test conjectured quantum limits on transport as well as relations between transport and quantum chaos. A larger bond dimension may be needed to capture quantum entanglement over the growing thermal coherence length, making these calculations more demanding. 

Besides transport and hydrodynamics, we showed that the method can be used to extract information on quantum chaos. Using a measure for the distance between reduced density matrices corresponding to the perturbed and unperturbed initial states we obtained the characteristics of the chaotic wave that precedes the emergence of hydrodynamics.
We found that the chaotic front propagates ballistically with a ``butterfly" velocity $v_B$, but at the same time broadens diffusively (i.e., $\delta x\propto \sqrt{t}$). Because the front broadens one cannot describe the chaos in terms of a single growth rate or Lyapunov exponent. Indeed if we try to define a local Lyapunov exponent then its value is seen to fall off with the distance from the perturbation.
The chaotic properties we found are consistent with recent analysis of an effective model for a spin chain evolved with random unitary gates \cite{VonKeyserlingk2017, Nahum2017}. It would be interesting to compare these results directly with a TDVP approximation of the time dependence  generated by the random unitaries as well as more realistic non conserving dynamics generated, for example, by a Floquet drive.

The rapid convergence of the chaos parameters with increasing bond dimension raises interesting questions concerning the crossover from classical to quantum chaos. For low bond dimension the TDVP constitutes nonlinear classical dynamics of almost local variables. The emergence of chaos is then governed by the nonlinearity. On the other hand at the largest possible bond dimension the TDVP gives the exact many-body quantum dynamics, which is linear. This can still be viewed as classical Hamiltonian dynamics but in an enormous space of highly non-local variables. In this limit the chaotic dynamics is a result of coarse graining this huge space. It is reasonable to expect that the thermal coherence length provides a natural coarse graining scale on which it is possible to describe quantum chaos using non-linear classical equations.

Another interesting application of the method is to study the many-body localization transition at infinite temperature by approaching it from the thermal regime. In this case the calculation will become more demanding because of the need to capture the  entanglement in the rare insulating inclusions (Griffith regions). The size of such regions, $l_G$, grows together with the diverging correlation length, $\xi$ as  $l_G \sim \xi \log(L/a)$ on approaching the MBL critical point \cite{Vosk2015,Potter2015}. 

Before closing we comment on two recent preprints posted after the appearance of the first version of our paper. In one paper, White et al. \cite{White2017} introduced a method to compute the time evolution of a mixed state (density matrix) using a new truncation scheme that retains exact information on local operators. An important difference from our method is that the density matrix truncation leads to irreversible time evolution. Another recent paper gave further justification to the method we present here based on the representability of thermal density matrices with convex combinations of matrix product states \cite{Berta2017}.

\section{Acknowledgements} 
This work has benefited from stimulating discussions with Erez Berg, Xiao Chen, Mark Fischer, Andrew Green, Markus Greiner, Curt von Keyserlingk, Roger Mong, Anatoli Polkovnikov, Shivaji Sondhi and Mike Zaletel.  We  thank Ashley Milsted for discussions on the implementation of TDVP and for sharing his code ``evoMPS" \cite{evoMPS}.
This research was supported in part by the ERC synergy grant UQUAM (EA and EL), Research Unit FOR 1807 through grants no. PO 1370/2-1, by the ERC starting grant QUANT MATT no. 679722, and the National Science Foundation under Grant No. NSF PHY-1125915. EA Acknowledges support from the Gyorgy Chair of Physics at the University of California Berkeley.

\newpage

\appendix
\section{Distance measures between two variational states}

Let $\ket{\psi_1(0)}$ be a representative state in the ensemble of initial conditions. This state is perturbed by a unitary operator localized at the left edge of the system to give $\ket{\psi_2(0)} = S^x_{0}\ket{\psi_1(0)}$.  We want to compare the time evolution of the perturbed state $\ket{\psi_2(t)}$ to that of the unperturbed state  $\ket{\psi_1(t)}$. To this end we compute a measure of distance  between the two reduced density matrices $\rho_1(x,t)$  and $\rho_2(x,t)$ that are  obtained from the above states by tracing out the first $x$ sites of the spin chain. 

The reduced density matrices can be written explicitly using the Schmidt decomposition of the states:
\bea
\ket{\psi_\a(t)}&=&\sum_{i=1}^{r(x,t)}\lambda_{\a, i}(x,t)\ket{\psi^L_{\a, i}(x,t)}\ket{\psi^R_{\a, i}(x,t)}\nn\\
\rho^R_\a(x,t)&=&\sum_{i=1}^{r(x,t)}\lambda_{\a, i}^2(x,t) \ket{\psi^R_{\a, i}(x,t)}\bra{\psi^R_{\a, i}(x,t)} \nn\\
&&
\eea
where $\a=1,2$. The superscripts $L,R$ refer to the left and right Schmidt states respectively. 
Here $r(x)$ is the Schmidt rank of the state  and $\lambda_{\a,i}(x)$ are the Schmidt values. If $\ket{\psi_\a}$ is taken to be an unconstrained random state then the Schmidt rank depends on the distance as $r(x)=min(2^x,2^{N-x})$. Note that the second number in the brackets, $D(x)\equiv 2^{N-x}$ is the dimension of the Hilbert space of the right partition. Thus we always have $D(x)\ge r(x)$.  If we constrain the evolution (e.g., by using TDVP) to matrix product states with bond dimension $\chi$ then $r(x)=min(2^x,2^{N-x},\chi)$.

We are now ready to estimate the Frobenius measure of the distance between the two reduced density matrices:
\be
d^2(x,t)=\text{tr}[(\rho^R_1(x,t)-\rho^R_2(x,t))^2].
\ee
Later we will also consider normalized distance measures. 

Because the initial states were chosen randomly they would generically have close to the highest possible entanglement, i.e., $\lambda_{\a i}(x,t)\approx 1/\sqrt{r(x,t)}$. The purity of the reduced density matrices with these Schmidt values is:
\be
\text{tr}[\rho^R_\a(x,t)^2]=\sum_{i=1}^{r(x)}\lambda(x,t)_{\a,i}^4\approx \sum_{i=1}^{r}{1\over r^2}\approx 1/r(x).
\label{purity}
\ee
In order to estimate the distance measure, we also need to assess the overlap between the two distinct density matrices:
\begin{widetext}
\be
\text{tr}[\rho^R_1(x,t)\rho^R_2(x,t)]=\sum_{i,j=1}^r\lambda_{1,i}^2\lambda_{2,j}^2|\langle\psi^R_{1,i}(x,t)|\psi^R_{2,j}(x,t)\rangle|^2\approx {1\over r(x)^2}\sum_{i,j=1}^r|\langle\psi^R_{1,i}(x,t)|\psi^R_{2,j}(x,t)\rangle|^2.
\ee
\end{widetext}
At  early time, before the perturbation can affect the right partition, the right Schmidt bases corresponding to the two states are identical and we must have $\langle\psi^R_{1,i}(x,t)|\psi^R_{2,j}(x,t)\rangle= \delta_{ij}$. Therefore, in early times the cross term is 1 and it exactly cancels the purities (\ref{purity}) to give $d^2(x,t)=0$. On the other hand, at very late times the perturbed and unperturbed states are expected to evolve into essentially uncorrelated random states in the Hilbert space. In this case the Schmidt states are also uncorrelated and we should have $|\langle\psi^R_{1,i}(x,t)|\psi^R_{2,j}(x,t)\rangle|^2\approx 1/D(x)$. The distance measure $d^2(x,t)$ is expected to approach
\be
d^2(x,\infty)\approx {2\over r(x)}-{2\over D(x)}.
\label{eq:dinf}
\ee
Note that as long as the right partition is larger than half the system, i.e. $x<N/2$, then $r(x)<D(x)=2^{N-x}$. This implies that the late time distance measure $d^2(x,\infty)$ is non vanishing in this case both for the exact evolution ($r(x)=2^x$) and for the variational time evolution (for which $r(x)\le\chi$). On the other hand, when the right partition is smaller than half the system, i.e. $x>N/2$, there is a dramatic difference between the variational and exact result. In the variational evolution $r(x)=\chi<D(x)$ also for $x>N/2$, leading to $d^2(x>N/2,\infty)>0$. Under the exact evolution, on the other hand, $r(x)=D(x)$ for $x>N/2$ and the above calculation yields a vanishing result for $d^2(x>N/2,\infty)$. This  can be understood physically as a consequence of thermalization. Under generic time evolution both states $\ket{\psi_1(t)}$ and $\ket{\psi_2(t)}$ become uncorrelated micro-states that represent the same thermal macro-state. They become indistinguishable to any sufficiently local measurement. The reduced density matrices of sub-regions smaller than half the system then approach the same thermal density matrix for both systems.

It is convenient to work with a normalized distance measure, which approaches unity at long times. One possible choice is the normalization used in the main text $\delta^2(x,t)=d^2(x,t)/d^2(x,\infty)$. Another sensible normalization is the following
\bea
{\underline{d}}^2(x,t)&=&{\text{tr}[(\rho^R_1(x,t)-\rho^R_2(x,t))^2] \over \text{tr}[\rho^R_1(x,t)^2]+\text{tr}[\rho^R_2(x,t)^2]}\nn\\
&=&1-{2\text{tr}\rho^R_1(x,t)\rho^R_2(x,t)\over \text{tr}[\rho^R_1(x,t)^2]+ \text{tr}[\rho^R_2(x,t)^2]}.
\eea
When the states become uncorrelated at long times, this measure of the distance between them approaches ${\underline{d}}^2(x,\infty)\approx1-r(x)/D(x)$. For the case of interest to us, i.e. $x<N/2$ the limiting value is therefore very close to unity because $D(x)\gg r(x)$ whether the Schmidt rank is restricted to to $\chi$ as in the variational approach or unrestricted as in the exact evolution. On the other hand, for $x>N/2$ the distance measure ${\underline{d}}^2(x,\infty)$ vanishes similarly to the unnormalized measure $d^2(x,\infty)$ if the system is undergoing the exact, or unrestricted evolution, while it is remains close to unity in the variational evolution, which restricts $r(x)<\chi$.

As mentioned above, we are interested in the case $x<N/2$. In this range the result of the TDVP calculations are found to be essentially the same for the two distance measures $\delta^2(x,t)$ and ${\underline{d}}^2(x,\infty)$.


\section{Comparison of TDVP  to the exact time evolution in a small system}

\begin{figure}[t]
\includegraphics[width=0.45\textwidth]{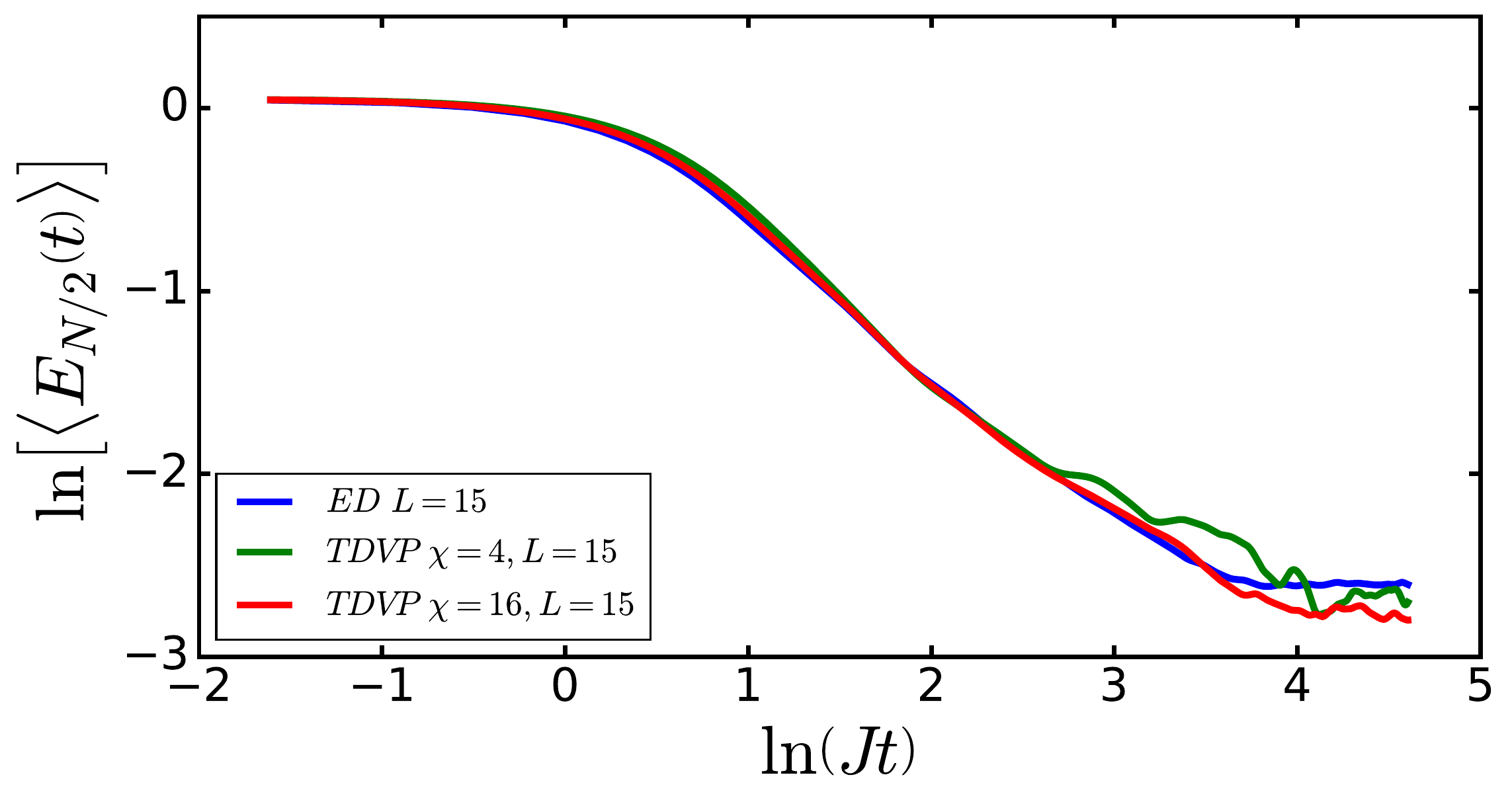}
\llap{\parbox[c]{13.5cm}{\vspace{-7.7cm}\textbf{(a)}}}
\includegraphics[width=0.45\textwidth]{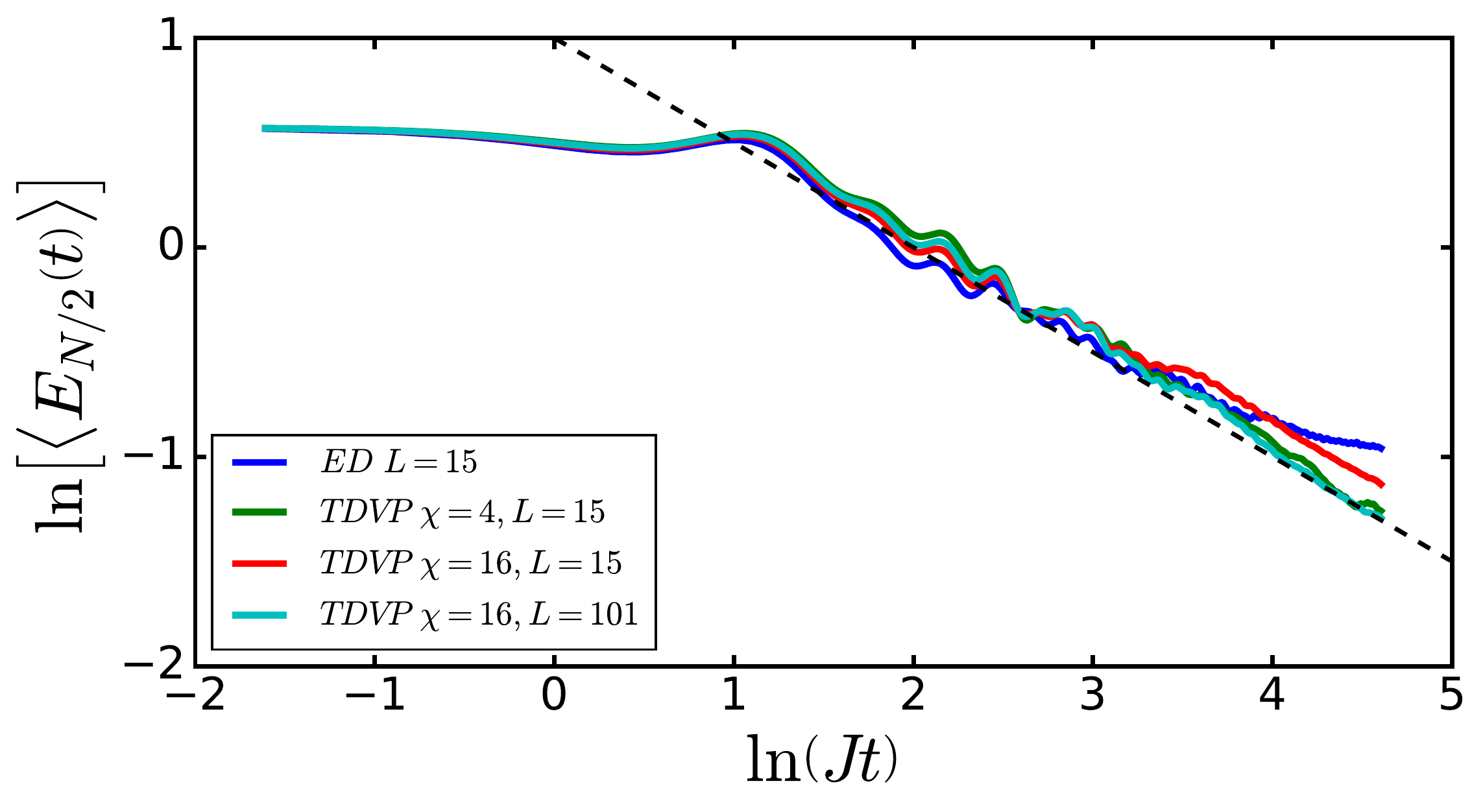}
\llap{\parbox[c]{13.5cm}{\vspace{-7.7cm}\textbf{(b)}}}
\caption{Comparison between TDVP and exact calculation for energy relaxation after a local quench in a small system of 15 sites. (a) The same interaction parameters as in the main text $\{J,h_x,h_z\}=\{4,1,-2.1\}$. (b) Modified interaction parameters to obtain faster emergence of diffusive relaxation: $\{J,h_x,h_z\}=\{4,4.5225,3.545\}$ } 
\label{fig:diff_app}
\end{figure}
\begin{figure}[t]
\includegraphics[width=0.45\textwidth]{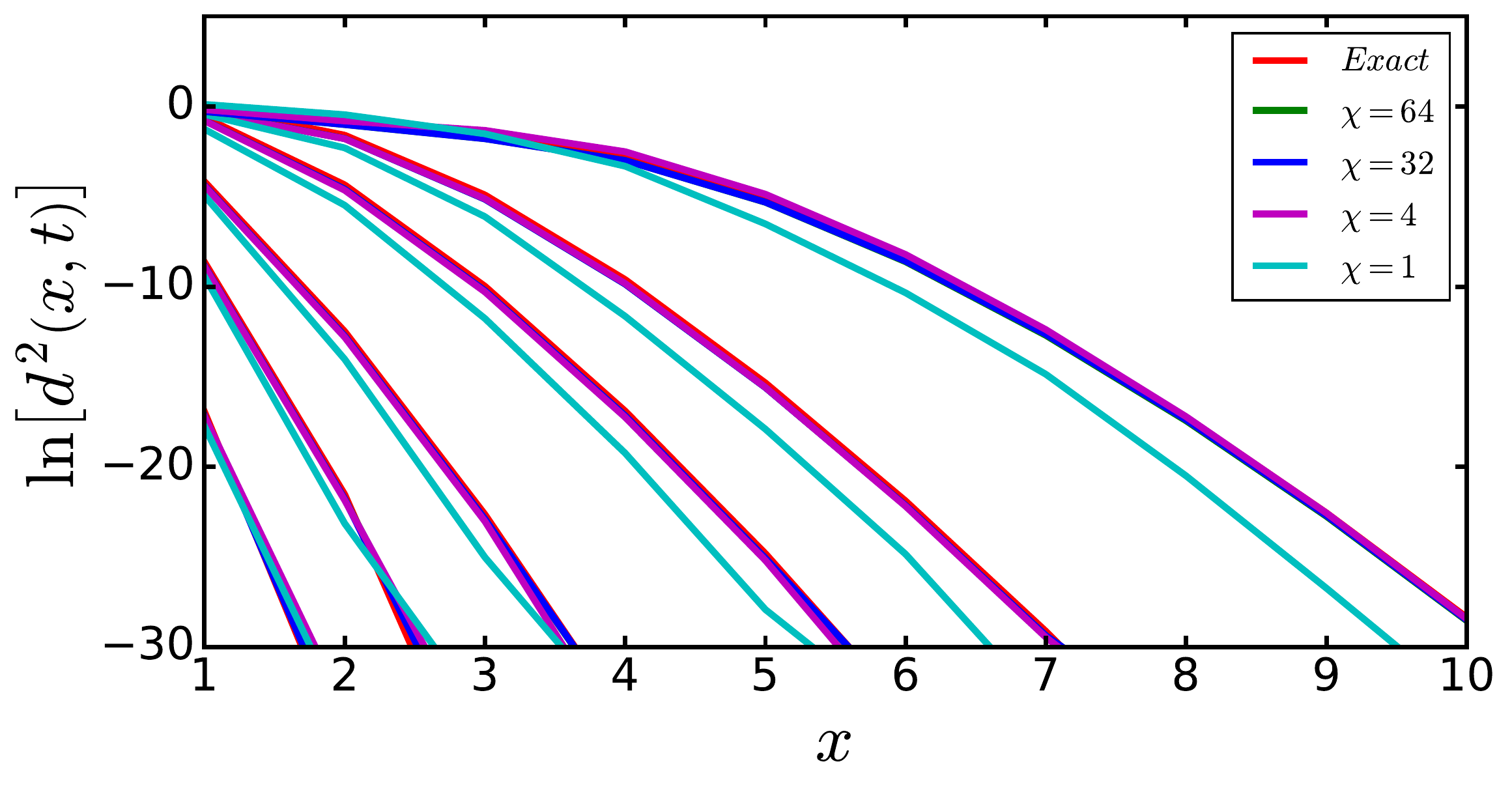}
\caption{Comparison between TDVP and exact calculation for the chaos indicator in a small system of 20 sites.
The un-normalized distance $\ln d^2(x,t)$ is plotted as a function of the distance from the perturbation, at times $Jt=0.04,0.32,1,3,5,10$. For each time we show the curves corresponding to TDVP calculations with varying $\chi$ and the exact time-evolution. } 
\label{fig.small_sys}
\end{figure}

\noindent In the main text we have shown convergence of the results with increasing bond dimension already at small values of $\chi\approx 4$. To further support this convergence we turn to a small system of 15 ($N=14$) or 20 ($N=19$) sites, where exact results are readily obtained. Below we compare the exact results to TDVP calculations with varying bond dimension $\chi$. We perform these tests for the two quench schemes considered above. First we consider relaxation of local observables following a non unitary perturbation $S^+_{N/2}$ applied in the middle of the chain. Second, we consider the propagation of the chaotic front in the small system after application of a unitary perturbation to the left edge of the system.

Fig.~\ref{fig:diff_app} shows the relaxation of the local energy density after the nonunitary perturbation is applied to the middle of a 15 site chain. For the parameters used in the main text of the paper this system is too small to reach the diffusive regime, as seen in panel (a) of that figure. Nevertheless, TDVP with $\chi=16$, which is much smaller than the full Hilbert space ($\chi=128$), shows very good agreement with the exact result (up to finite size effects). We can change the model parameters to the optimized values $\{J,h_x,h_z\}=\{4,4.5225,3.545\}$ in order to obtain diffusive relaxation already at shorter times, which can be seen even in a system with 15 sites. The exact result for the energy relaxation following a local quench in this system is shown in Fig.~\ref{fig:diff_app}(b). Again we see excellent convergence with the bond dimension $\chi$.

Turning to chaos, Fig.~\ref{fig.small_sys} shows $\ln d^2\left(x,t\right)$ (the un-normalized distance measure) calculated both with exact time-evolution and using TDVP with various values of $\chi$ (note that for 20 sites $\chi=1024$ is needed to get the exact time evolution). It is clear from these results that good convergence is  achieved already at $\chi\approx 4$. At larger values of $\chi$ the difference between the calculations for different bond dimension show up only in the saturation value $d^2(x,\infty)$, which is expected to behave as $d^2(N/2,\infty)\sim 1/\chi$ according to 
Eq. (\ref{eq:dinf}).

\section{Effective Lyapunov exponent}
It is natural to define the instantaneous  Lyapunov exponent as
\bea
\widetilde{\lambda}_L \left( x,t \right) \equiv \frac{\partial }{\partial t} \ln \delta\left( x,t \right).
\label{eq:lambdaL}
\eea
Fig.~\ref{fig.Lyapunov_x,t} shows this instantaneous Lyapunov exponent as a function of time for various distances $x$ from the perturbation. 

We are interested in the effective Lyapunov exponent, $\widetilde{\lambda}_L(x)$, defined as value of $\widetilde{\lambda}_L \left( x,t \right)$ at the time when $\delta^2$ reaches some threshold value above machine precision. In order to assess the dependence of $\widetilde{\lambda}_L(x)$  on $x$ we hypothesize a form for $\ln \delta$ consistent with the scaling results of Fig.~\ref{fig.chaos_t} :
\be
\ln \delta \sim \frac{v_Bt-x}{\sqrt{D_Bt}}.
\ee
Plugging this into the definition (\ref{eq:lambdaL}) gives:
\be
\widetilde{\lambda}_L \left( x,t \right)  \sim \frac{x+v_Bt}{\sqrt{D_B}t^{3/2}}.
\label{eq:lambdaLxt}
\ee

The scaling form also suggests that the time $t_0$ at which $\delta^2(x,t)$ exceeds a set cutoff grows with distance as $t_0\sim x/v_B$. Plugging this into (\ref{eq:lambdaLxt}) results in
\bea
\widetilde{\lambda}_L \left(x\right) \sim \frac{v_B^{3/2}}{\sqrt{D_B x}} \sim \frac{1}{\sqrt{x}},
\eea
which is not too far from the numerical result shown in Fig. \ref{fig.chaos_t}

\begin{figure}[t]
\includegraphics[width=0.45\textwidth]{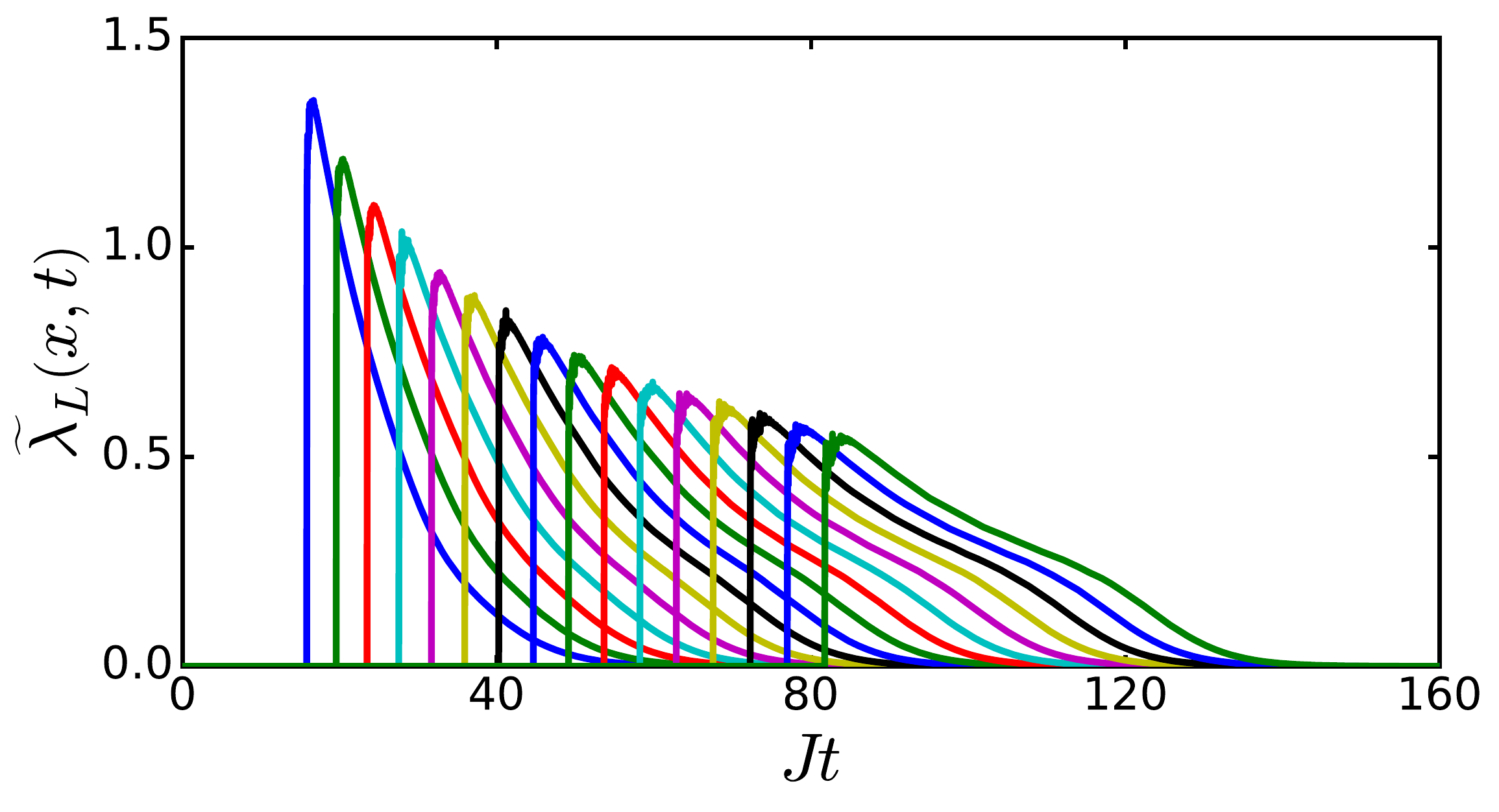}
\caption{The instantaneous Lyapunov exponent $\widetilde{\lambda}_L \left(x,t\right)$ as a function of time for distances $14\leq x \leq 44$ (in steps of $\Delta x=2$).}
\label{fig.Lyapunov_x,t}
\end{figure}

\end{document}